\documentclass[twocolumn,footinbib,floatfix,aps,10pt,prl]{revtex4-1}

\usepackage{graphicx}
\usepackage{epsfig}
\usepackage{epstopdf}
\usepackage{amsmath}
\usepackage{amsfonts}
\usepackage{txfonts}
\usepackage{amssymb}
\usepackage{amstext}
\usepackage[sort&compress]{natbib}
\usepackage{rotating}
\usepackage{dcolumn}
\usepackage{graphics}
\usepackage{array,multirow}
\usepackage[english]{babel}
\usepackage{attachfile}
\usepackage{pgfplots}
\usepackage[export]{adjustbox}
\usepackage{braket}

\def\be{\begin{equation}}
\def\ee{\end{equation}}
\def\bea{\begin{eqnarray}}
\def\eea{\end{eqnarray}}

\begin{document}

\title{Magnetic field enhancement of organic photovoltaic cells performance}

\author{S. Oviedo-Casado${}^{1}$, A. Urbina${}^{2}$, and J. Prior${}^{1}$}
\affiliation{${}^{1}$Departamento de F{\'i}sica Aplicada, Universidad Polit{\'e}cnica de Cartagena, Cartagena 30202 Spain\\
${}^{2}$Departamento de Electr{\'o}nica, Universidad Polit{\'e}cnica de Cartagena, Cartagena 30202 Spain}

\begin{abstract}
Charge separation is a critical process for achieving high efficiencies in organic photovoltaic cells. The initial tightly bound excitonic electron-hole pair has to 
dissociate fast enough in order to avoid photocurrent generation and thus power conversion efficiency loss via geminate recombination. Such process takes place assisted by transitional states that 
lie between the initial exciton and the free charge state. Due to spin conservation rules these intermediate charge transfer states typically have singlet character. Here we propose a 
donor-acceptor model for a generic organic photovoltaic cell in which the process of charge separation is modulated by a magnetic field which tunes the energy levels. The impact of a magnetic field is 
to intensify the generation of charge transfer states with triplet character via inter-system crossing. As the ground state of the system has singlet character, triplet states are 
recombination-protected, thus leading to a higher probability of successful charge separation. Using the open quantum systems formalism we demonstrate that not only the population of triplet charge 
transfer states grows in the presence of a magnetic field, but also how the power outcome of an organic photovoltaic cell is in that way increased. 
\end{abstract}

\maketitle

\section{Introduction}

\begin{figure}[!t]
\includegraphics[width=8cm]{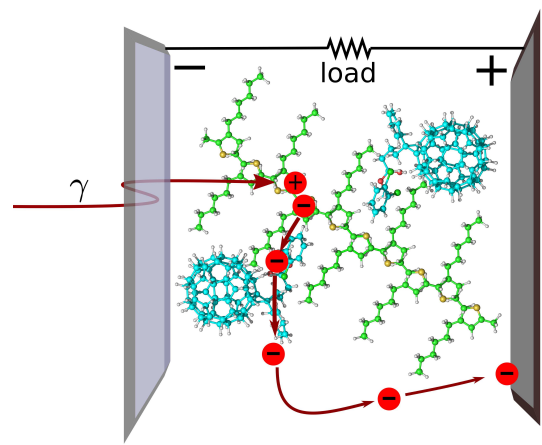}
\caption{Simplified portrayal of a bulk nanostructure heterojunction P3HT(Poly(3-hexylthiophene-2,5-diyl)):PCBM solar cell, where incoming photons induce electronic transitions to a higher state of a 
$\pi$ orbital in a monomer of --in this case-- P3HT. Thus a Coulomb bound electron-hole pair is formed, from which the electron rapidly migrates to the interface of the polymer with the 
acceptor material (fullerene derivative PCBM here), whereupon charge separation takes place and a net photocurrent is created, leading ultimately to electricity generation.}
\label{Figure1}
\end{figure}

Despite presenting near unity absorbed photon-to-electron quantum efficiencies in a broad range of incident photon wavelengths, organic photovoltaic donor-acceptor (D-A) cells have overall power 
conversion efficiencies that do not surpass 11\% \cite{EfficiencyTable,Dimitrov2014,Cao2012,Scharber2006}. Among the most efficient organic photovoltaic (OPV) cells to date are those based on bulk  
heterojunctions composed of 
conjugated polymers blended with fullerene derivatives, such as the one presented in Fig. [\ref{Figure1}] \cite{Sheng2015,Park2009,Vandewal2009}. Numerous approaches have focused on the study of the 
impact that morphology and interfaces of the blend have in the charge generation and charge transfer processes \cite{Botiz2014,Vandewal2013}. Another line of research emphasizes the role of charge 
transfer states at the distributed interface of the donor and acceptor moieties of the blend. Very detailed studies link the relative position of the energy levels of bands and trapped states to the 
global recombination rates and therefore to the photogeneration and final power conversion efficiency \cite{Deibel2010,Deotare2015}. The role of disorder --which creates defects and traps-- has been 
pointed out as one of the main factors which  diminishes the potential for high power conversion efficiency from the external quantum efficiency values to the much lower power conversion efficiency 
that is finally obtained. On the other hand, it has also been demonstrated that an overly crystalline structure with very low disorder is a drawback in efficiency \cite{Noriega2013}. Fine-tuning the 
position of the energy levels will hence provide a tool to manipulate the charge transfer and recombination traits of a given cell. Accordingly, the OPVs design strategy has to be tackled as a 
trade-off among D-A domain sizes, wave function delocalisation, diffusion length, and disorder in order to maximise charge separation and therefore power conversion efficiency.

OPVs photogeneration dynamics initial stages are nowadays so optimised that both photon absorption and exciton dissociation take place in the femtosecond time-scale ($\approx$ 100), and with high 
quantum efficiency \cite{Bakulin2012,Gelinas2014a}; it is thus in the subsequent evolution where efforts have to focus. Importantly, notice that exciton dissociation happens with the hole remaining 
spatially fixed in the donor. The process then has to be understood as the excited electron travelling to the polymer-fullerene interface, where long-range delocalisation of the electron wave function 
in eigenstates of the fullerene acceptor allows to avoid rapid recombination and enables ultrafast charge separation \cite{Gelinas2014a}. Since recently, free charge (FC) state formation --i.e. 
complete electron-hole dissociation-- is thought to happen in small steps or ``jumps'' via intermediate charge transfer (CT) states coupled with strong environment vibrational modes 
\cite{Bakulin2012,Chin2014,Tamura2013,Prior2015}. Charge transfer states, also known as polaron pairs, are weakly bound intermolecular e-h pairs whose role is to mediate charge generation in organic 
photovoltaic devices \cite{Veldman2009}. Lying at the D-A interface, CT properties such as delocalisation (size) much depend on its optical characteristics as well as on their mutual 
coupling. They in turn determine the maximum voltage attainable at the OPV, thus becoming critical if the aim is to improve the performance \cite{Shockley1961,Vandewal2009,Deotare2015}. However, due 
mainly to their experimental inaccessibility (they are mostly dark states, with very low dipole strength \cite{Vandewal2009}), and the wide diversity of structures and materials a general model is 
lacking, making the comprehension and manipulation of these states both difficult and compelling. It is therefore in the charge transfer stage of photocurrent generation where the challenge and 
opportunities to improve the efficiency of OPVs abide. 

The ground state of a conjugate polymer has zero spin. Consequently an electron which jumps to an excited state in a perturbative regime (such as by absorption of a solar photon) has to conserve the 
spin, and will form a singlet (total spin S = 0) exciton conditioning all subsequent evolution \cite{Barford2005}. In particular, it means that the CT state to which the e-h pair evolves will also be 
a singlet. The formation of triplet CT states happens only via inter-system crossing (ISC) from the singlet CT, or when a free electron is trapped by the Coulomb well of a hole in a process 
denominated 
non-geminate recombination, which renders triplet and singlet CTs in a 3:1 ratio. Dissociation from CT to free charge states is equally probable regardless of the CT state spin, whereas geminate 
recombination to the ground state is spin dependent. While singlet CT recombines in a time scale much similar to the FC formation rate (i.e nanoseconds), a triplet CT cannot directly recombine to a 
singlet ground state, thereby the only recombination pathway available is decaying to a triplet exciton which can then undergo triplet-triplet annihilation. Whenever acceptor structure allows for 
enough delocalisation of CT states, CT to triplet exciton is suppressed \cite{Rao2013a}, thus favouring FC formation from triplet CTs. Hence, the importance of spin dynamics is highlighted in these 
systems. It is in the light of the recent discoveries regarding spin in OPVs \cite{Chang2015,Rao2013a}, and drawing from ideas about singlet-triplet radical pair reactions in biological physics 
\cite{Schulten1978, Thorsten2004}, that we propose here that magnetic field can increase the ISC, thereby enhancing triplet CT states formation and leading to a higher performance exhibited in the 
form of a higher generated photocurrent.


This article is organised as follows. We first motivate the presence of a magnetic field by accounting for its possible effects, focusing on its impact in charge transfer states. We then 
introduce a generic model for a conjugate polymer-fullerene derivative photovoltaic cell that contains all the essential characteristics, and proceed to discuss the microscopic details and the 
influence of the magnetic field, showing the potential it has to enhance free charge generation. In the last section, we calculate the increase in photocurrent intensity and power that the magnetic 
field induces, thus demonstrating its capacity as a tool for exploring photocurrent generation dynamics.


\section{Inter-system crossing and magnetic field effects on electron-hole pairs}

ISC --i.e. the transformation between singlet and triplet spin states-- occurs spontaneously via spin-orbit interactions and the overlap of singlet and triplet states due to delocalisation by coupling 
with environment vibrations \cite{Kohler2009}. Both effects depend on the spatial separation of the pair, decreasing exponentially with e-h separation. Accordingly, for charge transfer states the 
combined effect will not be larger than a few meV even at the closest distances \cite{Cohen2009,Zhang2012}. The only other natural source of singlet to triplet conversion is the hyperfine coupling of 
the spins with the local magnetic environment created by the atomic nuclei surrounding the spin particle. This effect is typically much smaller (10$^{-2}$-10$^{-4}$ meV \cite{Cohen2009}) but 
distance independent and it will consequently dominate when spatial separation is big. 

Applying an external magnetic field causes the three spin states from the triplet CT to split due to the Zeeman effect, the splitting being proportional to the magnetic field as 
$2\mu_BgS_zB$ with $\mu_B$ the Bohr magneton and $g$ the Land\'e factor. Thus the $S_z$ = 1 state of the triplet will lie closer to the singlet CT, hence favouring resonant ISC transitions. However, 
it has also been pointed out that this may cause the $S_z$ = 1 state to be higher in energy than the singlet, thereby forbidding ISC transitions \cite{Chang2015}. Accordingly, OPV design should 
target to accomplish small spatial diffusion of CT states to maximise singlet-triplet energy separation, hence increasing the positive impact a magnetic field without going out of resonance. Due to 
the Zeeman splitting, it could also be argued that the higher triplet CT state might be made resonant not with the singlet CT but with the FC state,  greatly increasing charge dissociation. 
Yet given typical CT-FC energy separations of tens of meV 
\cite{Chang2015}, this resonance is only achieved with magnetic fields stronger than 100 Tesla, when typical experimental set-ups would typically reach 35 T at most. A third approach is nevertheless 
conceivable: the split spin states may become resonant with intermediate vibronic (namely electronic-vibrational mixed) states that 
lie in between the charge transfer and free charge states. This possibility would also have the potential to improve the OPVs performance and the set-up would require just to fine-tune the magnetic 
field until resonance is fulfilled. 

Aside from Zeeman splitting, an external magnetic field adds up to the internal magnetic field created by the atomic nuclei, increasing in that way the inter-system crossing via hyperfine 
interaction. In this case the relevant Hamiltonian is
\be
\mathcal{H}_{hyf} = -\mu_B\Bigl(g_e(B + a_eI_e)\cdot S_e + g_h(B + a_hI_h)\cdot S_h\Bigr),
\ee
where $B$ is the external magnetic field, \emph{a} are the hyperfine coupling constants and I are the hyperfine magnetic fields \cite{Schulten1978,Frankevich1992}. Both terms cause each of the spins 
to precess with a frequency $g\mu_BB$ in such a way that when it happens in a randomly oriented spin environment such as that of hyperfine coupling, it induces spin flips that lead to transitions from 
the singlet to the triplet state, and backwards \cite{Cohen2009,Hontz2015}. 

Previous studies of the effect that a magnetic field has in organic solar cells are based on long time ($\mu$s) OPV dynamic models, with mostly negative magnetic field effects in photocurrent 
generation 
\cite{Deotare2015,Hontz2015}. The reason being that in such time-scale singlet and triplet CT states are completely degenerate, and transitions to triplet $\pm 1$ spin states are Zeeman splitting 
forbidden due to changes in the transition amplitudes caused by the differences in precession frequencies \cite{Hontz2015}. Crucially, our model differs radically in that we consider a much faster 
time-scale dynamics, as pointed out already in models like those by Friend et al. \cite{Rao2013a,Gelinas2014a}. In our approach, both CT dynamics and precession frequency are on the nanosecond 
time-scale, where singlet and triplet CTs are still energetically apart and the scheme we described above for ISC holds. In such a scenario, the spin states evolution is still permitted and,provided 
the 
triplet CT has a longer lifetime than the singlet, a magnetic field will enhance the generation of free charges \cite{Frankevich1992}.

In addition to increasing the conversion of spin states via inter-system crossing, a magnetic field also delays decay of triplet charge transfer states into triplet excitons \cite{Merrifield1971}. 
This adds up to the assumption that triplet recombination pathway can be already suppressed by design \cite{Rao2013a}. Thus the lifetime of triplet charge transfer states increases due to an external 
magnetic field, permitting in this way the formation of more free charges. The effect of the magnetic field on tightly bound excitons is negligible and will not be 
considered in our analysis \cite{Yan2009}.

\section{Donor-acceptor organic photovoltaic model in a magnetic field}

\begin{figure}[!t]
\includegraphics[width=7.5cm]{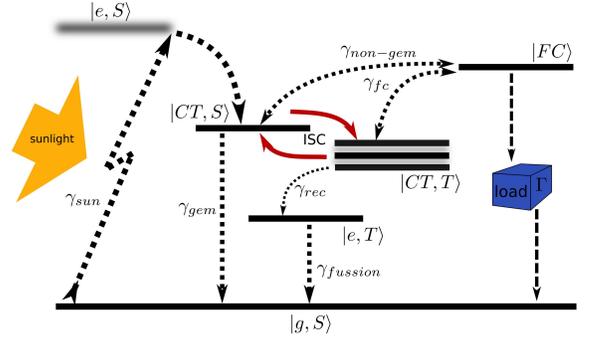}
\caption{
Schematic representation of the basic energy levels involved in the process of photocurrent generation in an OPV. Incoming photons create singlet exciton states, which rapidly evolve to become 
singlet charge transfer states. Inter-system crossing --represented here by solid arrows-- can transform these singlet states into triplet CTs, that are Zeeman split by an external magnetic field. 
From charge transfer states, either the e-h pair recombines to the ground state or dissociation takes place and a free charge is generated, in incoherent processes represented by dotted arrows whose 
relative strength represents the characteristic rates for the corresponding transitions. An abstract load between the free charge and the ground state allows to study the I-V characteristics of the 
model.}
\label{Figure2}
\end{figure}

Theoretical modelling of polymer based solar cells is usually involved, and approximations such as the unidimensional conjugate polymer character have to be used \cite{Mitchison2015a}. 
Nonetheless, the general framework for the process of photocurrent generation --represented in Figure [\ref{Figure1}] for a schematic P3HT:PCBM solar cell-- can be simplified to a handful of 
appropriately chosen states. Here we will consider the model as depicted in Fig. [\ref{Figure2}], whereupon initial excitation of the singlet ground state $\ket{g,S}$ (whose energy $E_0$ is for 
simplicity hereinafter taken as zero) an exciton state is formed. However, the exciton is very short lived, and within 100 fs the electron is separated from the hole and a charge transfer state 
with singlet character $\ket{CT,S}$ is formed \cite{Gelinas2014a}. As singlet CT state formation happens in a much faster time scale than the subsequent evolution (which is in the nanosecond 
range), and its quantum efficiency is nearly one; we will assume that excitation occurs directly to the singlet $CT$. 

The Hamiltonian describing the system in Fig. [\ref{Figure2}] reads
\be
\begin{split}
\mathcal{H}_S &= E_0\ket{g,S}\bra{g,S} + E_s\ket{CT,S}\bra{CT,S} \\& + E_t\ket{CT,T}\bra{CT,T}  + E_{et}\ket{e,T}\bra{e,T} \\&  + E_{FC}\ket{FC}\bra{FC} 
 + \mathcal{H}_m,
\end{split}
\label{Hamiltonian}
\ee
containing all the relevant energy levels. Namely Eq. [\ref{Hamiltonian}] features the ground state, a singlet charge transfer state, a triplet charge transfer state, in which a sum of three 
degenerate 
spin states is implicitly assumed, a triplet character exciton (i.e. the tightly bound form of the triplet CT) and a free charge state to represent the dissociated electron-hole pair state. In 
addition, S = 0 and T = 0,$\pm$1 indicate the singlet or triplet character of the state. Likewise 
\be
\begin{split}
\mathcal{H}_m &= \sum_{T=0,\pm 1}\Bigl(2\mu_B g T B\ket{CT,T}\bra{CT,T}\\& + ISC(\ket{CT,S}\bra{CT,T} + \ket{CT,T}\bra{CT,S})\Bigr),
\end{split}
\ee
describes the Zeeman splitting caused by an external magnetic field, as well as the coherent singlet-triplet CT interchange. Here, ISC contains all the magnetic field contributions to the coherent 
singlet-triplet CT interchange that we described in the previous section.

Besides unitary evolution, there are possible transitions among the energy levels which due to their stochastic nature cannot be accounted for in a Hamiltonian evolution. These are the dissociation 
and recombination of charges, that depend upon the specific structural characteristics of the polymer-fullerene blend and the 
temperature. As we are dealing with coherent spin evolution and incoherent states transitions, and provided that the interesting dynamics lies well outside the time evolution domain where non-secular 
 and non-Markovian effects are of relevance \cite{Oviedo2016}; we choose the formalism of Lindblad-type master equation evolution for open quantum systems \cite{BreuerPetruccione}, in which each 
transition is considered via a non-Hermitian operator of the form 
\be
\mathcal{L}_{\alpha} = \gamma_\alpha \left[\sigma_\alpha \rho(t) \sigma_\alpha^\dagger - \frac{1}{2} \left\{ \sigma_\alpha^\dagger \sigma_\alpha,\rho(t) \right\}\right],
\label{Lindblad}
\ee
where each $\alpha$ accounts for a different incoherent transition in Fig [\ref{Figure2}] specified here by the transition operators $\sigma_\alpha$ and with relaxation rates $\gamma_\alpha$. 

\begin{figure}[!t]
\includegraphics[width=\columnwidth]{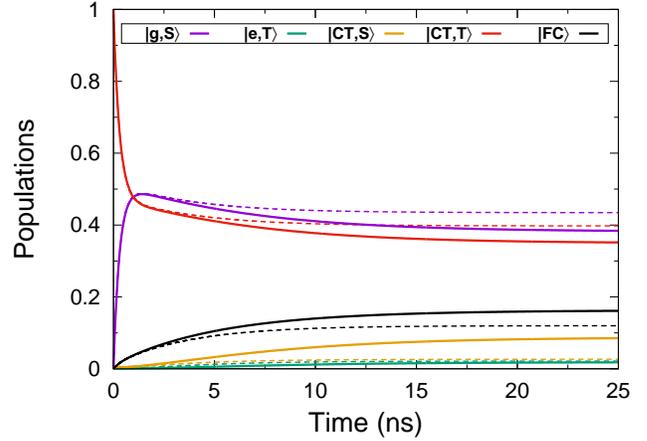}
\caption{Time evolution of the model energy levels populations in the absence (dashed) and presence (continuous) of a 25 T magnetic field. We have chosen a small triplet CT to triplet exciton decay 
rate of 2 ns to enhance the action of the magnetic field for displaying purposes. We will show in the next section how a faster rate also produces a noticeable photocurrent increase in the presence 
of a magnetic field. Geminate recombination and free charge generation both have a 1 ns rate, while non-geminate recombination is 1.5 ns and triplet fusion is 0.5 ns. Both triplet CT and FC 
populations at the steady state increase in the presence of a magnetic field, as hypothesised.}
\label{Figure3}
\end{figure}

Figure [\ref{Figure3}] shows the numerical time evolution of energy level populations from the Hamiltonian Eq. [\ref{Hamiltonian}] both in the presence and absence of an external magnetic field of 25 
T, according to the Lindblad master equation
\be
\frac{d\rho}{dt} = -i\left[\mathcal{H}_S,\rho(t)\right] + \sum_\alpha \mathcal{L}_\alpha \left[\rho(t)\right],
\ee
where complete positivity (i.e. physical) evolution is guaranteed \cite{BreuerPetruccione}. The diagonal elements of the density matrix correspond to the population of the different energy levels 
shown in Fig. [\ref{Figure2}]. Notice the increase in the free charge energy level population at the steady state when in the presence of a magnetic field. Such increment translates into a net 
increase in photocurrent as we will demonstrate in the next section. In addition, triplet charge transfer state population is also augmented at the expense of singlet charge transfer, pointing out as 
well at reduction in geminate recombination, as evidenced by the decrease of population in the ground state. 

Results shown are robust against most of the decay rates' variations. Nonetheless, notice that as was already mentioned in \cite{Rao2013a}, whenever the decay path through triplet-triplet fusion 
grows 
sufficiently fast together with rapid triplet charge transfer state conversion to triplet exciton, the magnetic field shows a negative effect, as in such situation generating more triplets 
essentially induces faster recombination to the ground state. However, whenever we are at room temperature, and following the simple design recipe demonstrated in \cite{Rao2013a} of having a well 
ordered acceptor should suffice to avoid this problem and turn the presence of a magnetic field into an advantage for OPVs performance, as well as a useful instrument to investigate the energy 
levels structure of the different solar cells. 

\section{From quantum transfer rates to I-V curves.}

The performance of a solar cell can be studied through its intensity-voltage (I-V) characteristic, as well as from the power that can be extracted defined as $P = I\cdot$V. To analyse the 
behaviour of our model under a magnetic field as was presented in the previous section, we use the framework of quantum heat engines \cite{Shockley1961,Alicki1979}. Such 
formalism has been already successfully applied to a wealth of quantum systems \cite{chin2013role,Jaemin2015}, in particular in quantum photosynthesis, unveiling the positive effect of quantum 
coherences on the performance of 
these systems \cite{Creatore2013,Dorfman2013,Killoran2015}. 

The quantum heat engine performance is calculated as the rate at which useful energy is transferred across a certain energy gap. For a solar cell, this rate is measured as the number of free charges 
that decay to 
the ground state through a load that acts as the impedance $\Gamma$ of the circuit, as shown in Fig. [\ref{Figure2}]. This scheme mimics that of an OPV in which the free electrons arrive at an 
electrode 
while the opposite electrode supplies the OPV with the missing electrons (see Fig. [\ref{Figure1}]). Thus the intensity with which current circulates is defined as $I = e\rho_{FC,FC}\Gamma$, with 
$\rho_{FC,FC}$ the population of the free charge state and $\Gamma$ the impedance acting via a Lindblad term as was defined in Eq. [\ref{Lindblad}]. The voltage is given by $V = E_{FC} - E_g + 
Tln\frac{\rho_{FC,FC}}{\rho_{gg}}$, with T the temperature of the environment which for the purposes of this article is always room temperature. The aim is to increase the power extracted from the 
solar cell. As the maximum attainable voltage (open circuit $V_{OC}$) is fundamentally limited for a given OPV \cite{Shockley1961}, the strategy has to be focused on increasing the current across 
the load, for which the more population in the FC state we create, the higher the power outcome will be. 

The fundamental difference with the results presented in the previous section is that to realistically simulate a solar cell, we have to consider a solar-like excitation on an initial state with the 
population in the ground state of the Hamiltonian Eq. [\ref{Hamiltonian}]. A ``sun'' energy source can be simulated considering appropriate statistically chosen parameters; in particular we follow 
earlier works and take the average number of photons to be $n_h$ = 60 000 with a corresponding transition rate $\gamma_{sun}$ = 0.1 \cite{Creatore2013,Dorfman2013,Killoran2015}; which will induce 
incoherent transitions from the ground state to the singlet CT state. Notice that we choose again to bypass the singlet exciton stage, as we explained in the previous section. Furthermore, measuring 
circuit intensity and voltage is done including another incoherent rate $\Gamma$ that is varied to simulate different circuit impedances.  

\begin{figure}[!t]
    \includegraphics[width=\columnwidth]{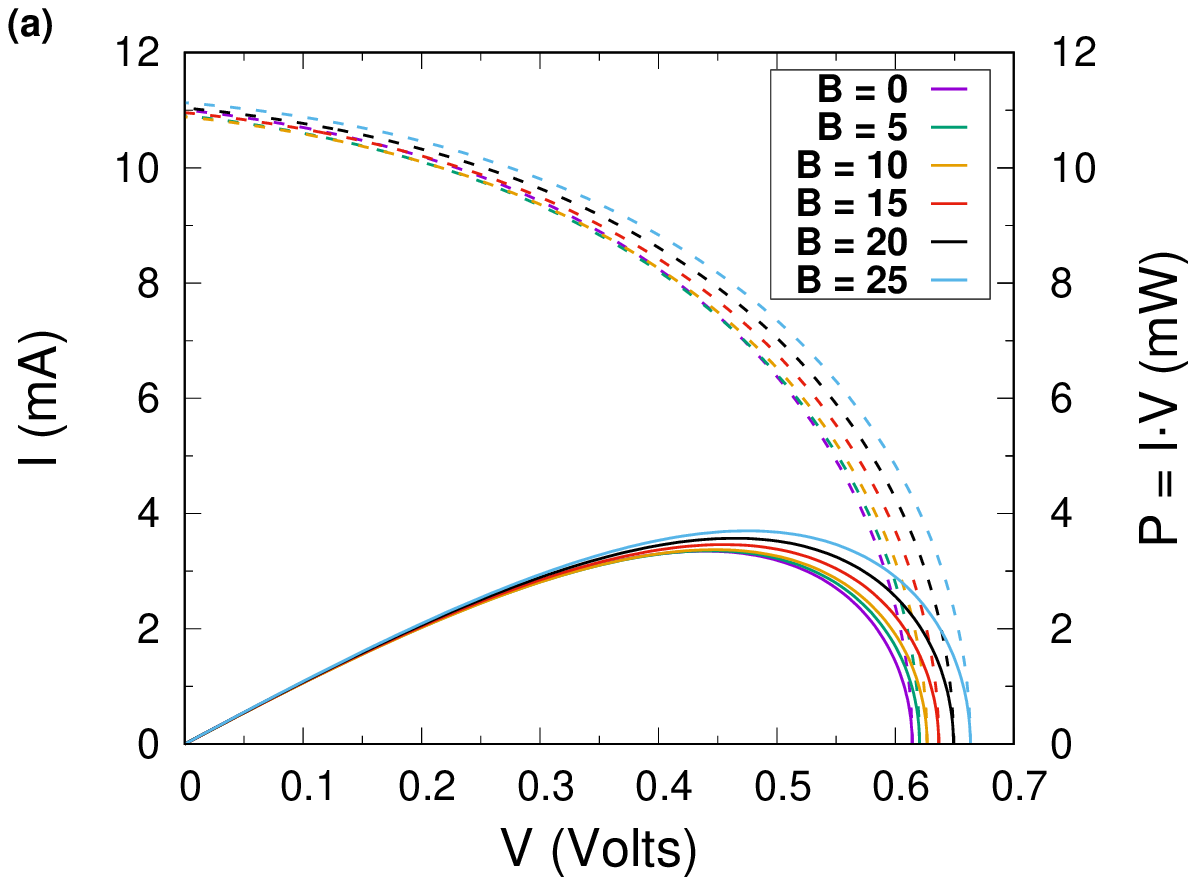}
    \includegraphics[width=\columnwidth]{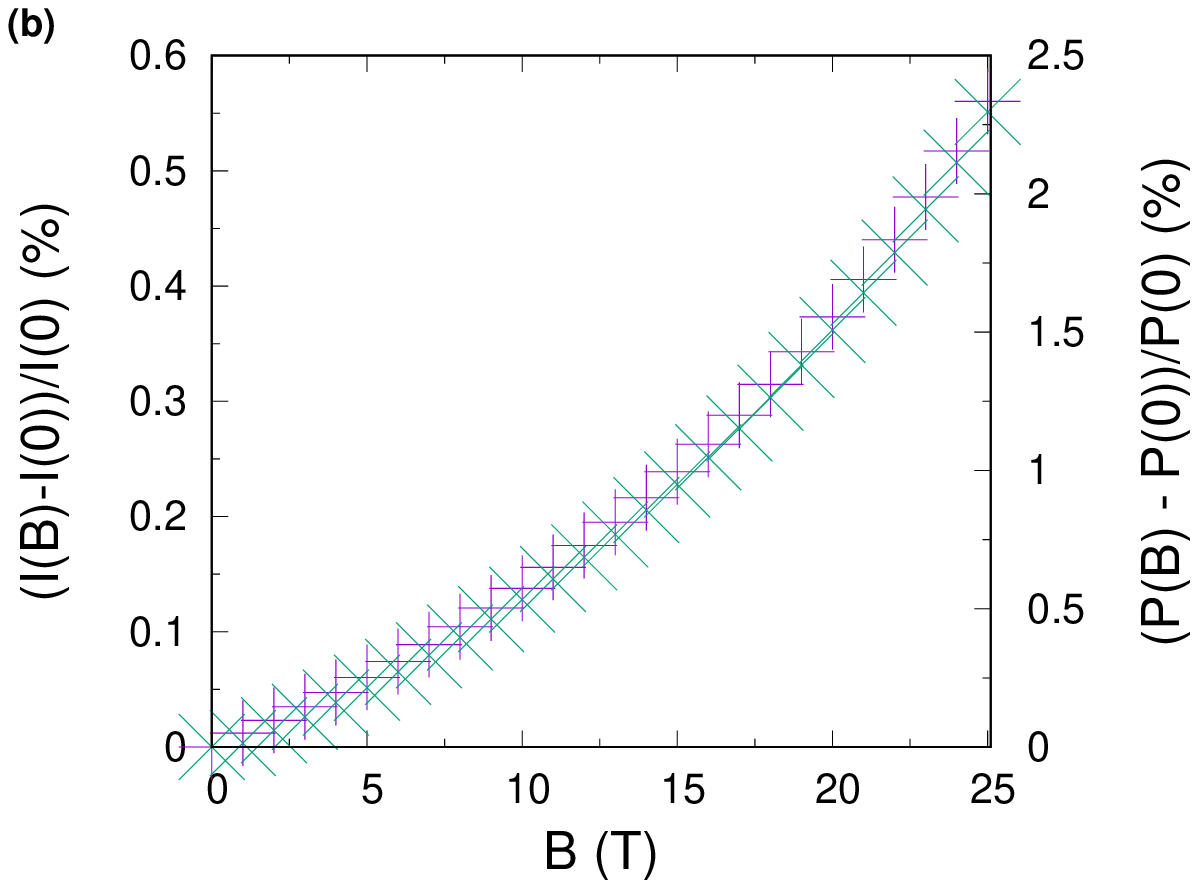}    
    \caption{Numerical I-V characteristic (dashed) and power (continuous) of the model solar cell shown in Fig. [\ref{Figure2}] for different magnitudes of the external magnetic field in (a). 
Percentage increase in the generated photocurrent across the load calculated for the point of maximum power ({\large{+}}) and percentage increase of the maximum power outcome ({\large{x}}) in (b). 
Geminate recombination and free charge generation both have a 1 ns rate, while non-geminate recombination is 1.5 ns and triplet fusion is 0.5 ns. Triplet exciton formation rate from triplet charge 
transfer states is here increased to 1.5 nanoseconds. It is demonstrated how the presence of a magnetic field enhances the performance of an organic solar cell.}
    \label{Figure4}
\end{figure}

Figure [\ref{Figure4}](a) shows the I-V characteristic for our model OPV in the presence of an increasing external static magnetic field together with the power generated. It is thus demonstrated the 
enhancing capacity of the magnetic field. At high impedance the population of the free charge state decays too fast and continuous excitation due 
to solar photons to the singlet CT state is in this case faster than the free charge formation due to the magnetic field enhancement of triplet CT, explaining why in this regime only the 
highest magnetic fields still have a positive effect on the intensity of the photocurrent generated. In particular, if impedances were increased further all curves would eventually unify. Intensity 
units are given as mA rather than also in intensity per unit area (mA/cm$^2$) owing to the rather arbitrary nature of the solar-like excitation we are employing. Hence it is shown how a magnetic 
field increases the capacity of the solar cell to create photocurrent.

Observe in addition that although it serves well the purpose of showing a magnetic field dependent photocurrent enhancement, the I-V characteristic  presented in Fig. [\ref{Figure4}](a) corresponds 
to a rather poor solar cell, as evidenced by the small filling factor (FF) that would correspond to the IV curves. Such parameter is defined through the maximum power as $P_{mp} = I_{mp}\cdot V_{mp} 
= FF\cdot I_{SC}\cdot V_{OC}$, where $I_{SC}$ is the current at zero voltage, and represents an independent measure of the corresponding OPV performance. The best filling factors are achieved solar 
cells whose I-V curves show a much more pronounced ``shoulder'', with an almost flat intensity for most voltages and an exponential drop when approaching $V_{OC}$. Our solar cell modelling presents 
several 
energy levels within the energy gap, which is equivalent to considering a diode with an ideality factor higher than unity, thus mimicking a realistic "non-ideal" solar cell. This is different from 
the equivalent circuit approach where adding a large series resistance and a small shunt resistance to the cell model also produces I-V curves with poor filling factor.

To more explicitly visualise the enhancing capacity of the magnetic field, in Figure [\ref{Figure4}](b) we represent the percentage variation in the intensity with the magnetic field, calculated at 
maximum power point, and correspondingly the percentage increase in maximum power. Not surprisingly we obtain again that introducing a magnetic field creates a positive effect in both the 
intensity and the power of the solar cell. These results show that it is possible to modify the photocurrent by introducing a magnetic field, thus validating our hypotheses which points to magnetic 
fields as a useful tool to tune the energy levels of triplet CT states and therefore explore the dynamics of photocurrent generation.

\section{Conclusions}

We have shown that a magnetic field can be a valuable experimental tool that provides a fine tuning of energy levels (triplets) relative to CT and other energy states relevant 
in the photocurrent generation. Methodical exploration of materials and blends for organic solar cells working under magnetic fields will allow the researchers for a fast screening of the optimum 
combinations regarding the relative energy level positions. Moreover, given the complicated experimental accessibility of certain critical states of organic solar cells, magnetic fields could be key 
in gaining a better comprehension of the inner structure and workings of OPVs.  

In addition, we have demonstrated the potential that a magnetic field has for enhancing the performance of polymer-fullerene compounds. By systematically exploring different sets of parameters it 
should be possible to find the optimal design protocols that make the presence of a magnetic field maximise photocurrent generation. Notice that the open circuit voltage increases in the presence of 
a magnetic field. The maximum open circuit voltage is determined by the Shockley-Queisser limit and occurs when the charges only recombine radiatively. Here we show that this limit can be overcome 
and that careful engineering of the process of CT formation at the polymer-fullerene interface could lead to improved devices. 

Finally, we have employed an open quantum system framework, in which we combine quantum coherent Hamiltonian evolution with the Lindblad master equation formalism. That way we are able to treat both 
the energy levels dynamics of the solar cell, quantum in nature, together with the influence that the polymer-fullerene structure has in the OPV performance, which is stochastic and has to be treated 
incoherently. While the open quantum systems formalism has been widely employed in a plethora of systems, it has only timidly been used to study polymer solar cells. Ours is then another step towards 
developing better tools to address and improve this systems.

\section{Acknowledgement}
Numerical calculations were performed using the QuTiP python package \cite{Qutip}. We acknowledge Alex W. Chin for useful discussions during the early stages of this article, Mark 
Mitchison for illuminating comments in the article preparation, and Luis Pedro Garc{\'i}a for his support with the cluster simulations. This research was undertaken with financial support from MINECO 
(SPAIN), including FEDER funds: FIS2015-69512-R together with Fundaci{\'o}n S{\'e}neca (Murcia, Spain) Projects No. 19882/GERM/15 and No. ENE2016-79282-C5-5-R.

\bibliography{OPVs}

\end{document}